# Signatures of superconducting and pseudogap phases in ultrafast transient reflectivity of Ca(Fe$_{0.927}$Co$_{0.073}$)$_2$As$_2$

SUNIL KUMAR[1(a)], L. HARNAGEA[2], S. WURMEHL[2], B. BUECHNER[2] and A. K. SOOD[1(b)]

[1] *Department of Physics, Indian Institute of Science, Bangalore 560012, India*
[2] *Leibniz-Institute for Solid State and materials Research, Dresden, D-01171 Dresden, Germany*

We present femtosecond time-resolved pump-probe spectroscopic studies of a pseudogap (PG) along with the superconducting (SC) gap in an overdoped iron pnictide Ca(Fe$_{0.927}$Co$_{0.073}$)$_2$As$_2$. It is seen that the temperature evolution of the photoexcited quasiparticle (QP) relaxation dynamics, coherently excited A$_{1g}$-symmetric optical phonon and two acoustic phonon dynamics behave anomalously in the vicinity of the superconducting transition temperature T$_c$. A continuous change in the sign of the experimentally measured transient differential reflectivity ΔR/R signal at the zero time-delay between the pump and probe pulses at a temperature of ~200 K is inferred as an evidence of the emergence of the PG phase around that temperature. This behavior is independent of the pump photon energy and occurs for crystals without the spin density wave phase transition.

Iron pnictides are the new generation high-T$_c$ superconductors which have many similarities with the cuprate superconductors, but have metallic behavior in the normal state rather than the Mott-insulator phase in cuprates [1]. Discovery of these iron containing superconductors have made the understanding of high-T$_c$ superconductivity even more challenging. Many experimental and theoretical studies have been reported in the last few years on new observational aspects about the role played by electron-phonon coupling, the spin and orbital fluctuations in the origin of high-T$_c$ superconductivity and the nature of electron- and hole-bands in the Fermi surface which develop a gap below the superconducting transition temperature T$_c$ [2-4]. A pseudogap phase has already been established in the case of cuprates [5-9]. However such a feature in the iron pnictides has not been established yet, though a few recent experiments have indicated that a PG phase needs to be invoked to explain the experimental data in the normal state, such as, ultrafast photo-induced reflectivity in SmFeAsO$_{0.8}$F$_{0.2}$ [10] and Ba(Fe$_{1-x}$Co$_x$)$_2$As$_2$ [11], NMR line-shapes in LaFeAsO$_{0.89}$F$_{0.11}$[12], Ba(Fe$_{1-x}$Co$_x$)$_2$As$_2$ [13], and Ca(Fe$_{1-x}$Co$_x$)$_2$As$_2$ [14], and infrared optical conductivity of both the electron and hole-doped BaFe$_2$As$_2$ [15-17]. It is expected that a knowledge of the unusual electronic properties in the normal state will be helpful in unraveling the origin of high-T$_c$ superconductivity.

In this paper, we show that femtosecond time-resolved reflectivity of an overdoped iron pnictide Ca(Fe$_{0.927}$Co$_{0.073}$)$_2$As$_2$ displays an evidence of a PG phase at a temperature of ~200 K where the non-equilibrium transient reflectivity shows a sign reversal from negative to positive with the decreasing temperature near the zero time-delay between the pump and the probe pulses. We have employed 395 nm (3.15 eV) femtosecond pump-pulses to prepare the system far from equilibrium and subsequently monitored its evolution in time by using 790 nm (1.57 eV) probe-pulses. Degenerate pump-probe experiments at 790 nm on another nearly optimally doped sample, Ca(Fe$_{0.935}$Co$_{0.065}$)$_2$As$_2$ confirm that the main features in the transient differential reflectivity signals are independent of the pump photon energy. In particular, the feature of sign reversal near zero delay between the pump and probe pulses, occurs in crystals which do not show the spin density wave phase transition. We have extracted the QP dynamics, and the coherent optical and acoustic phonon dynamics from the experimentally measured transient differential reflectivity of the sample. It is shown that the temperature evolution of the coherent optical and acoustic phonon dynamics behaves anomalously in the vicinity of the superconducting transition temperature T$_c$. The fastest exponential relaxation component in the recovery dynamics of the photoinduced reflectivity changes sign at a temperature of ~200 K. The sign of the time-resolved differential reflectivity ΔR/R in the pump-probe measurements is not unique and can depend strongly on the excitation power, photon energy or the orientation of laser polarization with respect to the crystal axis. It is noted that in a previous ultrafast pump-probe optical measurement on Ba(Fe,Co)$_2$As$_2$ [11], the observed sign of the ΔR/R was reported to be either positive or negative depending on the angle of the probe-polarization. Such anisotropy with respect to the probe-polarization was thought to be arising from a preferential ordering of the orthorhombic twin domains in the probed volume due to the anisotropic surface strain [11]. However, our observation of the sign reversal at a temperature of ~200 K is observed at fixed excitation power, photon energy, laser polarization and the probed region on the sample surface. This observation made using unfocussed pump and probe beams (large beam diameters of ~1.5 mm on the sample surface) is a clear indication of the intrinsic changes in the underlying electronic states topology leading to the pseudo-gap phase in the normal state and the superconducting phase at lower temperatures.

Single crystals of $Ca(Fe_{0.927}Co_{0.073})_2As_2$ used in our experiments have the superconducting transition temperature at $T_c \sim 15$ K, and do not show any magnetic transition as clear from the temperature-dependent resistivity and magnetization reported earlier in Ref. [18]. A Ti:Sapphire laser delivering ~45 fs pulses at ~790 nm and 1 kHz repetition rate was used. A larger fraction of the laser output was used to generate second harmonic pulses using a thin beta-barium-borate crystal. Unfocused pump pulses at 395 nm (beam diameter ~ 2 mm) and orthogonally polarized probe pulses at 790 nm (beam diameter ~ 1.5 mm) were used in our experiments to measure the transient differential reflectivity ΔR/R as a function of the time-delay between them in a non-collinear pump-probe geometry. The crystallographic orientations in our crystals are not known, however, they are cut along the z-axis and the pump and probe beams were kept at near normal incidence to the sample surface. Experiments were performed at sample temperatures varying from 3.5 K to room temperature with the sample mounted inside a continuous flow liquid-helium optical cryostat and at various pump-fluences while keeping the probe-fluence fixed at ~4 μJ/cm². During the entire set of experiments, the pump-probe beam polarizations and the angle of incidence as well as the probed region on the sample surface and the orientation of the crystal inside the cryostat were kept fixed.

In Fig. 1 we plot the experimental transient differential reflectivity ΔR/R data (thin black curves) measured at various sample temperatures T in Kelvin (K) as mentioned by using pump-fluence of ~32 μJ/cm². Here R is the nominal reflectivity value of the probe in absence of the pump and ΔR [= $R_{Pump\ on}$ - $R_{Pump\ off}$] is the change in the probe reflectivity induced by the pump. The experimental data at individual temperatures in Fig. 1 have been vertically shifted by a constant amount relative to the data at 3.5 K for better visibility of the temperature-dependence of various exponentially decaying components and the oscillatory contributions.

From Fig. 1, we note that the ΔR/R signal shows a continuous sign change from positive to negative at the zero-delay between the pump and probe at a temperature above ~200 K. The origin of such a behavior can be attributed to the emergence of a pseudogap-like phase around that temperature. The pseudogap phase has been invoked in understanding the temperature-dependence of the transient reflectivity of iron pnictides in a few recent ultrafast pump-probe studies [10,11] where the corresponding contribution in the QP dynamics is much faster and distinctly different from the contribution related to superconducting QPs at lower temperatures. In our present study, we see that the fast relaxation component that shows sign reversal at the $T_{PG} \sim 200$ K, is present at all the temperatures.

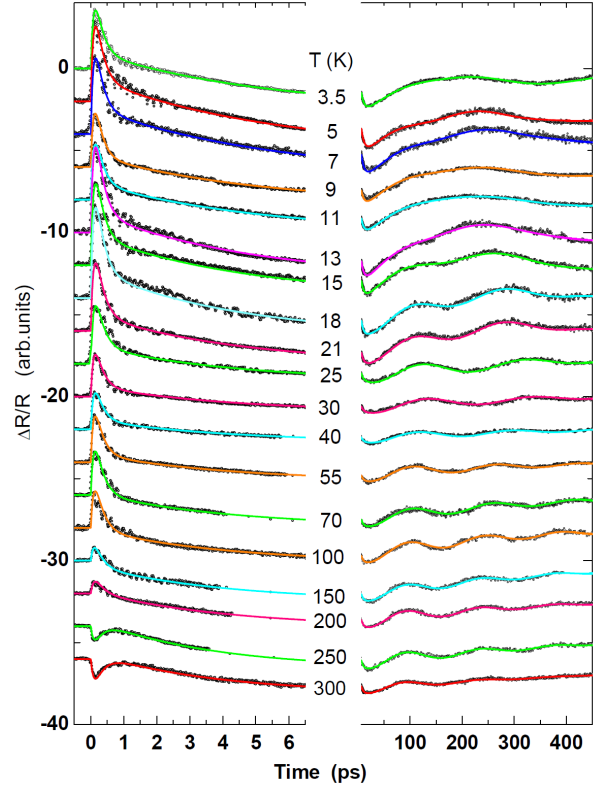

Fig. 1: (Color online) Time-resolved differential reflectivity data from $Ca(Fe_{0.927}Co_{0.073})_2As_2$ taken at various sample temperatures as mentioned. Solid lines of different colors are fits to obtain various exponentially decaying and oscillatory contributions.

Solid curves (colored) in the Fig. 1 are fits to the experimental data using a combination of three ($k$ = 1 to 3) exponentially decaying components $\sum_k A_k \exp(-t/\tau_k)$ having amplitude $A_k$ and corresponding time-constant $\tau_k$ attributed to the charged QP relaxation, and two ($l$ = 1 to 2) damped harmonic oscillatory contributions $\sum_l B_l e^{-t/\tau^p_l} \cos(2\pi \nu_l t + \phi_l)$ with amplitude $B_l$, dephasing time $\tau^p_l$, frequency $\nu_l$ and initial phase $\phi_l$ attributed to coherent acoustic phonon modes in the system.

At short time-scales of a few ps, the data also contains fast oscillations having an average time-period of ~300 fs. Coherent generation of a zone-center optical phonon by femtosecond pump-pulses is the origin of these fast modulations in the transient reflectivity. This particular mode is similar to previously reported observations [19,20] and is attributed to the $A_{1g}$ symmetric optical phonon related to As-atomic vibrations in the Fe-As planes of the pnictide lattice. It is conventional to analyze the optical phonon parameters in the Fourier domain. The Fourier analysis of the raw experimental data (Fig. 2a) taken up-to 6 ps is presented in Fig. 2b. To do this, the time-domain data is differentiated and 5-point averaged (inset of Fig. 2a) before the fast Fourier transform (FFT) is performed. The FFT at each sample temperature results into a Lorentzian mode as shown in Fig. 2b for T = 3.5 K. The continuous curve in Fig. 2b is a Lorentzian fit to the frequency domain data shown by open circles.

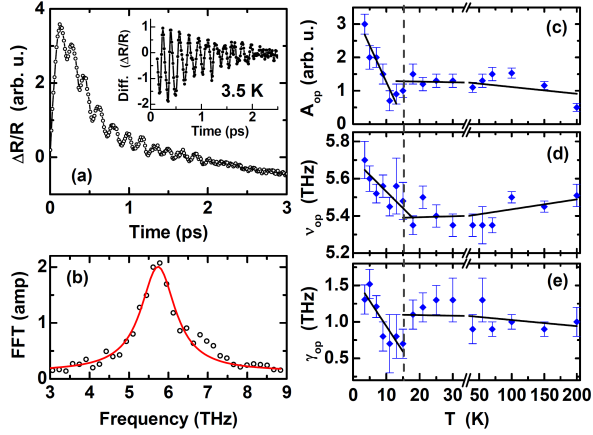

Fig. 2: (Color online) Coherent optical phonon dynamics. (a) Coherent oscillations at 3.5 K. The inset shows the differentiated data. (b) Fast Fourier transformation of the data in the inset of (a) along with Lorentzian fit (solid line). (c-e) Temperature-dependent amplitude $A_{op}$, central frequency $\nu_{op}$, and line-width $\gamma_{op}$ of the Lorentzian oscillator. Solid black lines in (c-e) are linear fits and the dotted vertical line have been drawn at nominal $T_c$.

The amplitude parameter $A_{op}$ in Fig. 2c which quantifies the phonon density of the coherently generated phonon mode following the photo-excitation, increases sharply below $T_c$. At the same time the phonon line-width $\gamma_{op}$ also shows an anomalous increase of ~100% as temperature decreases below $T_c$. The phonon frequency $\nu_{op}$ decreases slightly from 300 K to $T_c$ and then increases by a large amount of ~7%. The large anomalous temperature-dependence of the phonon parameters in the superconducting state suggests that the $A_{1g}$ optical phonon mode is strongly coupled with the superconducting QPs [21] resulting in large renormalization of the phonon self-energy.

The two slow oscillatory contributions in the transient reflectivity (Fig. 1) with time-period of oscillations in the range of 200 ps are similar to our previously reported results on underdoped $Ca(Fe_{0.944}Co_{0.056})_2As_2$ crystal observed using degenerate pump-probe spectroscopy at 790 nm [19]. The lower frequency mode is attributed to the transverse acoustic (TA) and higher one to the longitudinal acoustic (LA) mode. The temperature-dependences of the time-domain parameters of the two acoustic phonon modes are presented in Fig. 3. Similar to the previous observations on the underdoped compound [19], the temperature-dependences of the acoustic phonon parameters: amplitude ($B_{LA,TA}$) and dephasing time $\tau^p_{LA,TA}$ are highly anomalous in the vicinity of $T_c$ (solid lines in Fig. 3 have been drawn as guide to the eyes), implying that the elastic properties are influenced by the superconducting phase transition. At this stage, theoretical understanding of these results is lacking.

The main highlight of this study is the clear observation of a pseudogap phase with characteristic energy scale of ~33 meV as inferred from the temperature-dependent QP dynamics. As clearly visible from the experimental data in Fig. 1, the polarity of the transient reflectivity near zero-delay changes from negative to positive as temperature is lowered below 200 K. This implies that the residual normal state contribution (negative amplitude) above 200 K is dominated by a new (positive amplitude) component due to emergence of the pseudogap state at 200 K. The temperature evolution of the three-component QP relaxation dynamics in the sample is described in Fig. 4 where the amplitudes $A_{1,2,3}$ and corresponding time-constants $\tau_{1,2,3}$ as a function of the sample temperature have been presented by open symbols. Here, the vertical dashed lines represent the nominal superconducting transition temperature $T_c$ ~ 15 K and the continuous curves are theoretical models as described below.

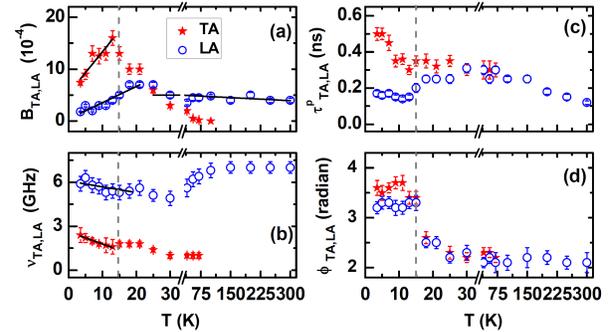

Fig. 3: (Color online) Temperature-dependence of the acoustic phonon parameters (amplitude B, frequency $\nu$, dephasing time $\tau^p$ and phase $\phi$) obtained for the LA and the TA modes. Thin solid are linear fits and the vertical dashed lines have been drawn at nominal $T_c$.

There is an important difference we would like to point out between our experiments and most of the previous experiments on similar systems, that we have used very low repetition rate lasers pulses which help us to use large diameter beams without focusing to a tiny spot on the sample and avoid the quasi-continuous heating of the sample [6]. Our results from fluence-dependent experiments reveal that the optical response is nearly linear upto pump-fluences of ~45 μJ/cm² for sample temperatures below the superconducting transition and linear for all fluences in the high-temperature normal state of the sample. In Fig. 5, the experimental data for the QP

Sunil Kumar *et al.*

dynamics with the fastest time-constant $\tau_1$ and three amplitudes $A_{1,2,3}$ of the exponentially decaying contributions in the transient differential reflectivity, taken at two sample temperatures of 9 K (in the superconducting state) and 70 K (in the normal state) are presented by open symbols and the dashed curves have been drawn as guide to eyes. The other two decay time-constants $\tau_2$ and $\tau_3$ are found to be pump-fluence independent. These results are similar to our previous results on underdoped $Ca(Fe_{0.944}Co_{0.056})_2As_2$ iron pnictide obtained using degenerate pump-probe experiments at 790 nm [22].

the doping level is reduced. This kind of behavior has been reported previously on 2-1-4 cuprate systems [8] where the effective pseudogap temperature was found continuously droping with the increasing doping level.

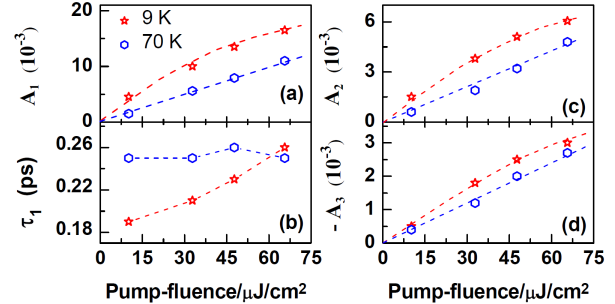

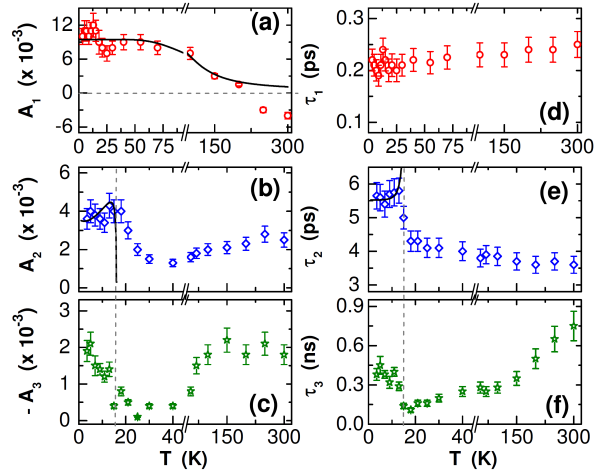

Fig. 4: (Color online) Temperature-dependence of the three exponential contributions with amplitude $A_{1,2,3}$ and decay-times $\tau_{1,2,3}$ in the transient reflectivity shown in Fig. 1. The vertical dotted lines represent the nominal $T_c$ and solid black lines are theoretical curves as described in the text.

Comparing our present results in Fig. 4 with those on underdoped $Ca(Fe_{1-x}Co_x)_2As_2$ (x = 0.056) system [22], we find that the trends in the temperature evolution are similar except for the reversal of sign of the amplitude $A_1$ from positive to negative below 200 K. Moreover, close inspection of the reported data in Fig. 2a of Ref. [22] indicate that the $\Delta R/R$ signal would show a sign reversal at the zero-delay at a temperature of ~350 K. Similarly, the undoped sample $CaFe_2As_2$ [23] is expected to reveal this type of behavior at even much higher temperature. In addition, recently reported studies using $^{75}As$ NQR measurements [14] on single crystals from the same series of $Ca(Fe_{1-x}Co_x)_2As_2$ iron pnictides had indicated that the pseudogap phase is confined to a tiny region between the spin density wave and the Fermi liquid phases in the x-T phase diagram and was predicted to vanish for the optimal doping and beyond. However, our present observations clearly suggest that the pseudogap phase does not vanish for the overdoping levels (x ~ 0.073) and comparisons can be drawn with previously reported results [22,23] leading to a conclusion that the pseudogap temperature increases as

Fig. 5: (Color online) Pump-fluence dependence of the three amplitudes $A_{1,2,3}$ and fastet time-constant $\tau_1$ shown for two sample temperatures of 9 K and 70 K. The dashed curves have been drawn as guide to the eyes.

We note from Fig. 4 that above 40 K, the values of $A_2$ and $A_3$ gradually increase to reach a maximum around the same temperature where $A_1$ switches sign from positive to negative. A maximum in the amplitudes of the QP relaxation components except the sign reversal of $A_1$, is similar to our previous results using 790 nm pump pulses. Furthermore, the QP relaxation dynamics in the $Ca(Fe_{1-x}Co_x)_2As_2$ systems is found to be different from previously reported other types of the iron pnictide systems [10,11,24-26]. We have found that irrespective of the photon-energy of excitation and the doping levels, the $Ca(Fe_{1-x}Co_x)_2As_2$ systems show three-component dynamics at all temperatures with large variations in the vicinity of the phase transition temperatures [22,23]. In comparison, systems from other iron pnictide families have shown one to three components [10,11,24-26]. Particularly, in the case of near optimally doped superconducting systems, like ours in the present study, previous reports have shown either a single component [24] or two components [25]. In the latter [25], these two components are associated with the recombination of QPs in the two hole-like bands *via* intraband and interband processes. Hence, the difference in the ultrafast optical response could be a consequence of the difference in their underlying low-energy electronic structures.

Quasiparticle creation across a charge-gap $\Delta$ can take place either by heating the sample or by optical excitation while the dynamics of the QP recombination across the gap is governed by phonon-creation and reabsorption [27]. At constant optical excitation-fluence ($\Phi$), the temperature-dependence of the photoinduced reflectivity provides the temperature-dependence of the photoexcited QP density and hence the recovery of the corresponding gap parameter $\Delta$. In the case of a temperature-independent gap $\Delta$, the temperature-dependence of the amplitude (A) of photoinduced reflectivity signal is proportional [27] to $\Phi\Delta^{-1}\{1+m\exp[-\Delta/k_BT]\}^{-1}$. Here, $k_B$ is the Boltzmann constant and m is a material-dependent constant given as $m = 2\nu/N(0)\hbar\Omega_c$, $\nu$ being the number of Boson modes per unit cell interacting with the quasiparticles, $N(0)$, the density of quasiparticle states at the Fermi energy and $\Omega_c$, the maximum cut off frequency of the Boson spectrum. For the other case where the gap is temperature-dependent, the amplitude is $\sim \Phi\{\Delta(T)+k_BT\}^{-1}\{1+m'\sqrt{k_BT/\Delta(T)}\exp[-\Delta(T)/k_BT]\}^{-1}$ and the corresponding relaxation time-constant is $\tau(T) \sim \Delta(T)^{-2}\log(1/\{g+\exp[-\Delta(T)/k_BT]\})$ with g being a fit parameter.

The complicated temperature dependences as seen in Fig. 4, makes it difficult to imply that different components in the low-energy spectrum can be distinguished by their different lifetimes [11]. However, since the lifetimes associated with the larger energy pseudogap $\Delta_{PG}$ are expected to be faster than those associated with the smaller energy SC gap $\Delta_C$, it is reasonable to attribute the fastest component $A_1, \tau_1$ (Figs. 4a, 4d) to the QP dynamics related to the pseudogap $\Delta_{PG}$ while the slow component $A_2, \tau_2$ (Figs. 4b, 4e) can be attributed to the QP recombination across the superconducting gap $\Delta_{C1}$. The third component having amplitude $A_3$ (Fig. 4c) and time constant $\tau_3$ (Fig. 4f) that is increasing linearly with the temperature in the normal state, could be associated with the QP coupling with low energy excitations such as acoustic phonons and/or spin-fluctuations [23,24] where variations in the temperature-dependences around $T_c$ are due to changes in the nature of the coupling when superconductivity is emerging.

Assuming an isotropic BCS-like temperature-dependent gap in the superconducting state as $\Delta_C(T) = \Delta_C(0)\{1-(T/T_c)^2\}$ where $\Delta_c(0)$ is the zero-temperature value of the gap, the above described model has been used to capture the behavior of $A_2$ and $\tau_2$ as shown by dark black curves in Figs. 4b and 4e providing an estimate of the zero-temperature value of the corresponding superconducting gap $\Delta_c(0)$. At the gap opening at the superconducting transition temperature $T_c$, the time-constant $\tau_2(T)$ shows quasi-divergence and has been quite well captured using the above described model for $\tau(T)$ (shown by dark line in Fig. 4e) using an appropriate value $\Delta_c(0) \sim 1.6k_BT_c$. For consistency, we use the same value of $\Delta_c(0)$ in the model function for the corresponding amplitude $A_2(T)$ and plot the result (dark line) in Fig. 4b which is quite off from the experimental results (symbols) near $T_c$ but matches well at low temperatures.

Ignoring the small feature seen in $A_1$ at $T_c$ (Figs. 4a and 4d), we can apply the same model as described above to capture the behavior of $A_1$ (data shown by symbols in Fig. 4a). Assuming that the quasiparticle dynamics below $T_{PG} \sim$ 200 K is governed by a temperature independent pseudogap, we have plotted the result from the RT-model as a dark line in Fig. 4a providing us an estimate of the vlaue of the pseudogap as $\Delta_{PG} \sim 1.9k_BT_{PG}$ and the parameter m $\sim$ 20. The model can best reproduce the experimental data only below 200 K and predicts values decresasing gradually to zero at temperature above 200 K.

Taking $\Omega_c = 0.1$ eV and $N(0) = 5$/eV/cell, the value of the number of Boson modes per unit cell participating in the quasiparticle recombination across the pseudogap is estimated to be $\nu \sim 5$. In the literature, various values of $\nu$ have been reported, e.g., $\nu \sim 4.4\pm1.5$ for Sm-1111 iron pnictide [10] or $\nu \sim 10$-$20$ for cuprates [7,8].

We note that the value of the superconducting gap extracted from the temperature-dependent QP dynamics is much smaller than the nominal BCS value and hence may indicate weak electron-phonon coupling in the iron pnictides, if phonon-mediated mechanism is responsible. On the other hand, the temperature-dependence of the $A_{1g}$ optical phonon mode as described in Fig. 2 and discussed earlier in the paper, counter-intuitively indicate a strong coupling of the superconducting QPs with this phonon mode. This discrepancy is not clear yet and a model description is beyond the scope of the present work. We hope that our results will motivate further theoretical studies for a better understanding.

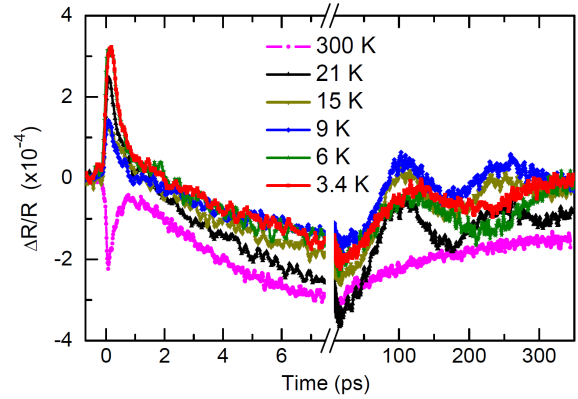

Fig. 6: (Color online) Time-resolved differential reflectivity data at various sample temperature and pump-fluence of $\sim$40 $\mu$J/cm$^2$, from near optimally doped Ca(Fe$_{0.935}$Co$_{0.065}$)$_2$As$_2$ crystal taken using 790 nm femtosecond pulses in the convential degenerate pump-probe experiment.

In the above, the experiental results have shown that the pump-probe technique is susceptible to changes in the low energy electronic structure. There are clear anomalies in the observables seen at the temperatures corresponding to the susperconducting gap and the pseudogap even when the optical photon energies of 1.57 eV and 3.15 eV are quite far from the underlying energy scales in the systems. To emphasize the fact that the main features in the transient differential reflectivity signals on our samples are independent of the pump photon energy, we show in Fig. 6, results from degenerate pump-probe experiments at 1.57 eV on another nearly optimally-doped sample, Ca(Fe$_{0.935}$Co$_{0.065}$)$_2$As$_2$. In particular, we can see that near the zero-delay between the pump and the probe pulses, the sign of the signal changes from positive at low temperatures to negative at room temperature.

Sunil Kumar *et al.*

Another important conclusion from our present and previously reported [22,23] experimental results on undoped, underdoped, nearly optimally doped and overdoped $Ca(Fe_{1-x}Co_x)_2As_2$ systems is that the initial (near zero time-delay) transient reflectivity changes sign at the optimal doping. This behavior is strikingly same as that observed in BSCCO cuprate systems [6] as a function of hole-doping level suggesting an abrupt transition in the quasiparticle dynamics at the optimal doping. Similar to the cuprate superconductors [7,8], we expect that the pseudogap exists for all doping levels in iron pnictides systems and the effective pseudogap temperature drops with the increasing doping level.

In summary, we have presented an experimental evidence of the high-temperature pseudogap state in superconducting $Ca(Fe_{0.927}Co_{0.073})_2As_2$ iron pnictide by measuring its temperature-dependent transient differential reflectivity. The estimated value of the pseudogap is ~33 meV. The superconducting gap with its zero-temperature value of $\sim 1.6 k_B T_c$ has been estimated by analyzing the temperature-dependence of the QP dynamics below $T_c$. In addition, a single coherent optical phonon mode at ~5.6 THz and two very low-frequency acoustic modes have been detected which have a similar dynamics as reported earliear in the case of underdoped Ca-122 system.

SK and AKS thank Department of Science and Technology, Government of India for financial support.

(a) Present address: *Department of Physics of Complex Systems, Weizmann Institute of Science, Rehovot 76100, Israel,*
E-mail: sunilvdabral@gmail.com
(b) asood@physics.iisc.ernet.in

[1] TESANOVIC Z., Physics **2** (2009) 60.
[2] STEWART G. R., Rev. Mod. Phys. **83** (2011) 1589.
[3] WEN H.-H. and LI S., Ann. Rev. Cond. Mat. Phys. **2** (2011) 121.
[4] JOHNSTON D. C., Advances in Physics **59** (2010) 803.
[5] VARMA C., Nature **468** (2010) 184.
[6] GEDIK N., LANGNER M., ORENSTEIN J., ONO S., ABE Y. and ANDO Y., Phys. Rev. Lett. **95** (2005) 117005.
[7] DEMSAR J., PODOBNIK B., KABANOV V. V., WOLF Th. and MIHAILOVIC D., Phys. Rev. Lett. **82** (1999) 4918.
[8] KUSAR P., DEMSAR J., MIHAILOVIC D. and SUGAI S., Phys. Rev. B **72** (2005) 014544.
[9] SCHNEIDER M. L., RAST S., ONELLION M., DEMSAR J., TAYLOR A. J., GLINKA Y., TOLK N. H., REN Y.H., LUPKE G., KLIMOV A., XU Y., SOBOLEWSKI R., SI W., ZENG X.H., SOUKIASSIAN A., XI X. X., ABRECHT M., ARIOSA D., PAVUNA D., KRAPF A., MANZKE R., PRINTZ J. O., WILLIAMSEN M.S., DOWNUM K.E., GUPTASARMA P. and BOZOVIC I., Eur. Phys. J. B **36** (2003) 327.
[10] MERTELJ T., KABANOV V.V., GADERMAIER C., ZHIGADLO N. D., KATRYCH S., KARPINSKI J. and MIHAILOVIC D., Phys. Rev. Lett. **102** (2009) 117002.
[11] STOJCHEVSKA L., MERTELJ T., CHU J.-H., FISHER I. R. and MIHAILOVIC D., Phys. Rev. B **86** (2012) 024519.
[12] AHILAN K., NING F. L., IMAI T., SEFET A. S., JIN R., McGUIRE M. A., SALES B. C. and MANDRUS D., Phys. Rev. B **78** (2008) 100501(R).
[13] NING F., AHILAN K., IMAI T., SEFET A. S., JIN R., McGUIRE M. A., SALES B. C. and MANDRUS D., J. Phys. Soc. Jpn. **78** (2009) 013711.
[14] BAEK S.-H., GRAFE H.-J., HARNAGEA L., SINGH S., WURMEHL S. and BUCHNER B., Phys. Rev. B **84** (2011) 094510.
[15] MOON S. J., SCHAFGANS A. A., KASAHARA S., SHIBAUCHI T., TERASHIMA T., MATSUDA Y., TANATAR M. A., PROZOROV R., THALER A., CANFIELD P. C., SEFET A. S., MANDRUS D. and BASOV D. N., Phys. Rev. Lett. **109** (2012) 027006.
[16] KWON Y. S., HONG J. B., JANG Y. R., OH H. J., SONG Y. Y., MIN B. H., IIZUKA T., KIMURA S.-i., BALATSKY A V., BANG Y., New J. Phys. **14** (2012) 063009.
[17] DAI Y. M., XU B., SHEN B., WEN H. H., HU J. P., QIU X. G. and LOBO R. P. S. M., Phys. Rev. B **86** (2012) 100501(R).
[18] HARNAGEA L., SINGH S., FRIEMEL G., LEPS N., BOMBOR D., HAFIEZ M. A., WOLTER A. U. B., HESS C., KLINGELER R., BEHR G., WURMEHL S. and BUCHNER B., Phys. Rev. B **83** (2011) 094523.
[19] KUMAR S., HARNAGEA L., WURMEHL S., BUECHNER B., SOOD A. K., Europhys. Lett. **100** (2012) 57007.
[20] MANSART B., BOSCHETTO D., SAVOIA A., ALBENQUE F. R-, FORGET A., COLSON D., ROUSSE A. and MARSI M., Phys. Rev. B **80** (2009) 172504.
[21] ZEYHER R., ZWICKNAGL G., Z. Phys. B **78** (1990) 175.
[22] KUMAR S., HARNAGEA L., WURMEHL S., BUECHNER B. and SOOD A. K., Solid State Communic. **160** (2013) 8.
[23] KUMAR S., HARNAGEA L., WURMEHL S., BUECHNER B., SOOD A. K., J. Phys. Soc. Jpn. **82** (2013) 044715.
[24] CHIA E. E. M., TALBAYEV D., ZHU J.-X., YUAN H. Q., PARK T., THOMPSON J. D., PANAGOPOULOS C., CHEN G. F., LUO J. L., WANG N. L. and TAYLOR A. J., Phys. Rev. Lett. **104** (2010) 027003.
[25] TORCHINSKY D. H., CHEN G. F., LUO J. L., WANG N. L. and GEDIK N., Phys. Rev. Lett. **105** (2010) 027005.
[26] GONG Y., LAI W., NOSACH T., LI L. J., CAO G. H., XU Z. A. and REN Y. H., New J. Phys. **12** (2010) 123003.
[27] KABANOV V. V., DEMSAR J., PODOBNIK B. and MIHAILOVIC D., Phys. Rev. B **59** (1999) 1497.